\documentclass[pra,showpacs,nofootinbib,superscriptaddress,12pt]{revtex4-1}
\usepackage{graphicx}% Include figure files
\usepackage{bm}% bold math
\usepackage{amssymb}
\usepackage{color}
\usepackage[usenames,dvipsnames]{xcolor}
\usepackage{subfigure}
\usepackage{amsmath}
\usepackage{amstext}
\usepackage{latexsym}
\usepackage[colorlinks=true,citecolor=Cerulean,linkcolor=Cerulean,urlcolor=Cerulean]{hyperref}
\usepackage{lipsum}
\usepackage{amsfonts}
\usepackage{epsfig}
\usepackage{dsfont}
\usepackage{mathrsfs}
\usepackage{arydshln,leftidx,mathtools}
\begin{document}
%\title{General unitarity constraints on the survival of a thermofield double state}
%\title{Quantum decay of the spectral form factor:\\ unitary constraints and exact results}
\title{Scrambling the spectral form factor:\\ unitarity constraints and exact results}

\author{A. del Campo}
\affiliation{Department of Physics, University of Massachusetts, Boston, MA 02125, USA}
\author{J. Molina-Vilaplana}
\affiliation{Universidad Polit\'ecnica de Cartagena. C/Dr Fleming S/N. 3
0202. Cartagena. Spain}
\author{J. Sonner}
\affiliation{Department  of  Theoretical  Physics,  University  of  Geneva,
24  quai  Ernest-Ansermet,  1211  Gen\`eve 4,  Switzerland}

\def\L{{\rm \hat{L}}}
\def\q{{\bf q}}
\def\l{\left}
\def\r{\right}
\def\te{\mbox{e}}
\def\d{{\rm d}}
\def\t{{\rm t}}
\def\K{{\rm K}}
\def\N{{\rm N}}
\def\H{{\rm H}}
\def\la{\langle}
\def\ra{\rangle}
\def\om{\omega}
\def\Om{\Omega}
\def\vep{\varepsilon}
\def\wh{\widehat}
\def\tr{{\rm Tr}}
\def\da{\dagger}
\def\iz{\left}
\def\zi{\right}
\newcommand{\beq}{\begin{equation}}
\newcommand{\eeq}{\end{equation}}
\newcommand{\beqa}{\begin{eqnarray}}
\newcommand{\eeqa}{\end{eqnarray}}
\newcommand{\intf}{\int_{-\infty}^\infty}
\newcommand{\into}{\int_0^\infty}
\newcommand{\cF}{{\cal F}}
\newcommand{\cI}{{\cal I}}

\begin{abstract}  

 Quantum speed limits set an upper bound to the rate at which a quantum system can evolve and as such can be used to analyze the  scrambling of information. To this end, we consider the survival probability of a thermofield double state under unitary time-evolution which is related to the analytic continuation of the partition function. We provide  an exponential lower bound to the survival probability with a rate governed by the inverse of the energy fluctuations of the initial state. Further, we elucidate universal features of the non-exponential behavior at short and long times of evolution that follow from the analytic properties of the survival probability and its Fourier transform, both for systems with a continuous and for systems with a discrete energy spectrum. We find the spectral form factor in a number of illustrative models, notably we obtain the exact answer in the Gaussian unitary ensemble for any $N$ with excellent agreement with recent numerical studies.  We also discuss the relationship of our findings to models of black hole information loss, such as the Sachdev-Ye-Kitaev model dual to AdS$_2$ as well as higher-dimensional versions of AdS/CFT.
 
\end{abstract}

%\pacs{
%67.85.-d %ultracold gases, trapped gases in quantum fluids and solids
%}
\maketitle

\section{Introduction}
Closed quantum systems undergo unitary dynamics, yet generically thermal behavior is observed at late times. This apparent contradiction is resolved by sharpening the notion of thermalization. If one considers observables associated with a subsystem and studies their nonequilibrium dynamics from an initial state that is not invariant under the system Hamiltonian, then after a thermalization time, the rest of the system acts effectively as a bath, rendering local observables thermal. The dynamics associated with interacting quantum systems leads to a redistribution of the information contained in the subsystem over the entire system, leading to a state with a high degree of entanglement between the subsystem and the rest. This process is often referred to as scrambling of information and it is particularly effective in chaotic systems. In fact, it has recently been proposed that the chaotic behavior leading to scrambling, can be characterized by a quantum version of a Lyapunov exponent\footnote{In fact this idea goes back to \cite{LarkinOvchi}, but has recently experienced a renaissance following the work of \cite{Maldacena15}.}, which obeys a strict bound \cite{Maldacena15}, defining quantitatively what we mean by a {\it maximally} chaotic quantum system.

Such a bound is also illuminating in the context of black hole physics. It has been argued that black holes in particular, are as effective at scrambling as physically admissible \cite{Hayden:2007cs,Sekino:2008he}, and therefore should saturate the aforementioned bound \cite{Maldacena15}. The connection to unitary quantum dynamics is made via the so-called AdS/CFT correspondence which states that certain gravitational theories are equivalent to unitary quantum field theories. One can thus rephrase the conjectures on fast scrambling of black holes in terms of unitarity constraints on the dynamics of (closed) quantum systems. A sharp way to formulate such constraints is by considering the late-time behavior of correlation functions, for example
\begin{equation}
G(t) = {\rm Tr}[ \rho{\cal O}(t) {\cal O}(0)], 
\end{equation}
where $\rho$ may be a thermal density matrix, $\rho \sim e^{-\beta H}$, or simply $\rho=|E\rangle \langle E|$ for some high-energy eigenstate. Despite the fact that black holes have finite Hilbert spaces, such correlations, or close variants thereof, typically decay to zero at late times if calculated in the semi-classical gravity approximation  \cite{Maldacena:2001kr}. Unitary quantum dynamics, however, implies that at late times the discrete nature of the spectrum forbids a continued decay and instead requires time averages of correlations do stabilize at $\sim e^{-{\rm S}}$, where ${\rm S}$ is the entropy of the thermal ensemble or simply the dimension of the associated Hilbert space. A particularly simple probe of the finiteness of the spectrum under unitary dynamics is the so-called spectral form factor, essentially the analytically continued partition function
\beqa\label{eq.SpecFormFac}
g(\beta,t)& = &|Z\left(\beta + it  \right)|^2 = \sum_{m,n}e^{-\beta(E_n + E_m)- it (E_n - E_m)},
\eeqa
where we have assumed that the spectrum is non-degenerate for simplicity. This function was the object of two recent studies with an eye on black-hole behavior, and in particular the issue of information loss. Firstly, in  the SYK \cite{Sachdev:1992fk,KitaevKITP} quantum mechanics, \cite{Cotler16,GarciaGarcia:2016mno} showed numerically that the initial decay eventually gives way to late-time random matrix behavior obeying unitarity bounds\footnote{Random matrix behavior has also been studied in a supersymmetric version of the SYK model \cite{Li:2017hdt}.}. Secondly, in 2D CFT with holographic duals, \cite{Dyer16} studied the evolution of $g(\beta,t)$ and estimated the onset of random-matrix behavior.
% An important subtlelty arises in models defined via an ensemble average, such as random matrix theory (RMT) or the SYK model. As we demonstrate below, the ensemble average and late-time limit do no commute. The ensemble average induces a kind of information loss, which leaves its imprint on the spectral form factor and similar quantities at late times. This may have important implications for the interpretation of the the SYK model, believed to be dual to an AdS$_2$ black hole.

Unitarity constraints on quantum dynamics can also come from a different angle, essentially by exploiting a version of the energy-time uncertainty relation \cite{MT45,Bhattacharyya83}. This results in universal bounds on the speed of quantum evolution, so called quantum speed limits (QSL). Such limits are often conveniently phrased in terms of the survival probability or fidelity of an initial state $| \psi_0\rangle$, to be defined below. In this paper we exploit the fact that the spectral form factor is closely related to the survival probability of the thermofield double state (TDS). We obtain bounds via QSL on the early time dynamics of this quantity and analyze the early and late-time dynamics of the TDS in a number of illustrative models. Consider a closed quantum system with Hilbert space ${\cal H}$. Then one may define a pure state in the tensor product ${\cal H}^{\otimes 2}$
\begin{equation}\label{eq.TFD}
| \psi(\beta) \rangle := \frac{1}{\sqrt{Z(\beta)}} \sum_ne^{-\beta\hat{H}/2}|n,n\ra\,,
\end{equation}
where $|n, n\ra$ denotes the tensor product of a given eigenstate $|n\ra$ in one copy with its CPT conjugate state $|\overline{n}\ra$ in the other.
 In this paper we examine the constraints posed by QSL on the survival probability of the TDS. In this case, because of its close connection to the partition function, we can express unitarity bounds on the {\it dynamics} of the system in terms of {\it equilibrium} thermodynamics.
 
 The paper is organized as follows.
 After establishing a relation between the analytic continuation of the partition function and the survival probability of a thermofield double state in Section \ref{sectionSP}, we elucidate various unitarity constraints  on the time evolution in Section \ref{sectionUC}. These include the long and short-time asymptotics of the survival probability as well as bounds on its decay derived from quantum speed limits. Section \ref{sectionEx} covers a selection of examples of growing complexity including the harmonic oscillator and the $xp$ model in AdS$_2$, the rational Calogero-Sutherland model, and the Gaussian Unitary Ensemble in random matrix theory. General features of the survival probability in AdS$_{d+1}$/CFT$_d$ are presented in section \ref{sectionAdS}, followed by a summary and discussion  \ref{sectionDis}.

\section{Partition function and  survival probability of a thermofield double state}\label{sectionSP}

Consider the TDS associated with a canonical state at inverse temperature $\beta$ (\ref{eq.TFD}).
When the dynamics is generated by a Hamiltonian of the form $\hat{H}_T=\hat{H}\otimes I_R$, the time evolution reads
\beqa
|\psi(\beta,t)\ra&=&\hat{U}(t,0)|\psi(\beta)\ra,\nonumber\\
&=&\frac{1}{\sqrt{Z(\beta)}}\sum_n e^{-(\beta/2+it)\hat{H}}|n,n\ra.
\eeqa
The survival amplitude is defined as the probability amplitude for the time-dependent state to be found in the initial state, i.e. as the overlap
\beqa
A(\beta,t)=\la\psi(\beta,0)|\psi(\beta,t)\ra=\frac{1}{Z(\beta)} \sum_ne^{-(\beta+it)E_n}.
\eeqa
The survival probability then reads
\beqa
\label{survateq}
S(\beta,t)&=&|\la\psi(\beta,0)|\psi(\beta,t)\ra|^2,\nonumber\\
&=&\frac{1}{Z(\beta)^2} \sum_{n,m}e^{-\beta(E_n+E_m)-it(E_n-E_m)}.
\eeqa 
This survival probability is evidently related to 
the analytic continuation of the partition function, and thus the spectral form factor introduced above (\ref{eq.SpecFormFac}).  We thus have
\beqa
S(\beta,t)&=&\left|\frac{Z(\beta+it)}{Z(\beta)}\right|^2.
\eeqa
As mentioned in the introduction this object has been proposed before \cite{BR03,BR04,PR15,Maldacena15,Dyer16,Cotler16} as a test of the spectral properties of black holes and as a measure of information loss. 
Our study relies on the fact 
that the survival probability is identical to the fidelity ${\cal F}$ between the initial quantum state $|\psi(0)\ra$ and  the corresponding time-evolving state $|\psi(t)\ra=\hat{U}(t,0)|\psi(0)\ra$,
\beqa
S(\beta,t)={\cal F}[|\psi(\beta,0)\ra,|\psi(\beta,t)\ra],
\eeqa
where the fidelity ${\cal F}$ between any two pure states $|\psi\ra$ and $|\phi\ra$ is defined as 
${\cal F}[|\psi\ra\la\psi|,|\phi\ra\la\phi|]=|\la \psi|\phi\ra|^2\in[0,1]$. A similar observation was made in the context of the boundary state quench in CFT by \cite{Cardy:2014rqa}. 
More generally, quantum states need not be pure, and given two density matrices $\rho_0$ and $\rho_t$ the fidelity reads 
%\beqa
${\cal F}\left( \rho_0,\rho_\tau\right)=\left[{\tr \sqrt{\sqrt{\rho_0}\,\rho_\tau\, \sqrt{\rho_0}}} \right]^2$ \cite{Uhlmann92}.
%\eeqa
As such, the survival probability is upper and lower bounded as
\beqa
0\leq S(\beta,t)\leq 1.
\eeqa
%Further, the evolution of $S(\beta,t)$ is constrained by QSL.
We will now proceed to derive further properties and bounds on $S(\beta, t)$ that follow from unitary time evolution in conjunction with certain properties of the spectrum of states $\rho(E)$. We will start by upper bounding the speed of evolution of the TDS in terms of {\it equilibrium} properties of the system.

\section{Constraints on the time evolution}\label{sectionUC}

\subsection{Quantum Speed Limits on the evolution of a thermofield state}

Quantum speed limits provide an upper bound to the rate of change of the survival probability, and more generally, to the fidelity between an initial density matrix $\rho_0$ and its time evolution $\rho_t$. The fidelity is useful to define a metric between quantum states in projective Hilbert space, known as the Bures length
\begin{equation}
\mathcal{L}\left(\rho_0,\rho_t\right)=\arccos{\left( \sqrt{{\cal F}\left(\rho_0,\rho_t\right)} \right) }\,
\end{equation}
that represents the angle swept out during time evolution as the state evolves from $\rho_0$ to $\rho_t$.
For unitary processes, two seminal results constrain the pace at which this happens. 
The Mandelstam-Tamm bound estimates the speed of evolution in terms of the energy dispersion of the initial state \cite{
MT45,Uhlmann92,Pfeifer93}, 
$\Delta E^2=\tr(\rho \hat{H}^2)-\tr(\rho \hat{H})^2$. 
Its original derivation relies on the Heisenberg uncertainty relation \cite{MT45}. 
The second seminal result is due to Margolus and Levitin \cite{ML98,LT09}, and provides an upper bound to the speed of evolution in term of the difference between the mean energy and the ground state energy, $E_0$.
Its original derivation relies on the study of the survival amplitude $A(t)=\la \psi(0)|\psi(t)\ra$. 
For dynamics generated by time-independent Hamiltonians, both bounds can be unified in the expression \cite{GLM03} 
\beqa
\label{QSL}
t\geq \tau_{QSL}\equiv \mathcal{L} \max\left(\frac{1}{E-E_0},\frac{1}{\Delta E}\right),
\eeqa
Equation (\ref{QSL}) provides a universal lower bound to the required time of evolution for the survival probability to decay by a Bures angle $\mathcal{L}=\arccos\sqrt{S(\beta,t)}$.

For a TDS the mean energy and its fluctuations are given by 
\beqa
 E&=& \la \psi(\beta)|\hat{H}|\psi(\beta)\ra=-\frac{d}{d\beta}\log Z(\beta),\\
\Delta E^2 
&=&\la \psi(\beta)|\hat{H}^2|\psi(\beta)\ra-\la \psi(\beta)|\hat{H}|\psi(\beta)\ra^2,\nonumber\\
&=&\frac{d^2}{d\beta^2}\log Z(\beta),
%&=& \frac{1}{Z(\beta)} \sum_ne^{-\beta E_n}E_n^2-\left(\frac{1}{Z(\beta)} \sum_ne^{-\beta E_n}E_n\right)^2,
\eeqa
that is, by the corresponding quantities in the canonical state at inverse temperature $\beta$.
Correspondingly energy fluctuations can be related to the specific heat of the system, defined as
\beqa
c_V=k_B\beta^2 \Delta E^2,
\eeqa
 in term of the Boltzmann constant $k_B$.

As was shown by Bhattacharyya \cite{Bhattacharyya83},  the Mandelstam-Tamm bound can be used to derive 
to lower bound the survival probability  by an exponential function.
For completeness we derive in Appendix \ref{GBB} a generalization of Bhattacharyya's bound that holds for arbitrary initial states. 
%[{\bf this is not needed as TFD is pure. of course, it's interesting, but maybe we should refer to it in a slightly different context?}] I agree, this seems to be the place .... I know. 
For the thermofield double state, this bound reads
\beqa
S(\beta,t)\geq \exp(-2\Delta Et),  
\eeqa
and holds for\ $S(\beta,t)\geq 1/2$ and $t\geq 0$ up to the half lifetime $t_h$ when $S(\beta,t_h)=1/2$.
As a result, the exponential bound to the survival probability can be rewritten as
 \beqa
\label{Fcv}
S(\beta,t)\geq \exp\left(-2\sqrt{\frac{c_V}{k_B}}\frac{t}{\beta }\right).
\eeqa
This is an important equation as it provides a strong connection between the quantum dynamics and equilibrium properties. 
Nonetheless, this bound is  restricted to a short time of evolution comparable to that governing the short-time asymptotic that we next discuss.

\subsection{Short-time nonexponential decay}
Under unitary evolution, the short-time decay of the survival probability is nonexponential \cite{FGR78}. 
Exploiting the power-series expansion of the time-evolution operator, the short-time asymptotic behavior of the survival probability reads 
\beqa
\label{stsurva}
S(\beta, t)=1-\left(\frac{t}{\tau_Z}\right)^2+\mathcal{O}(t^3), \quad \tau_Z\equiv\frac{1}{\Delta E}.
\eeqa
Hence,  at early times the energy fluctuations of the initial unstable state set the speed of evolution in Hilbert space.
The absence of a term linear in time in this expansion is a signature of unitary dynamics. Indeed, the nonunitary dynamics generated by an effective non-Hermitian Hamiltonian or induced by the coupling to a bath generally leads to the appearance of a correction $\mathcal{O}(t)$.
Yet, we also note that this short-time behavior relies on the existence of the second moment of the Hamiltonian and deviations are expected whenever energy fluctuations diverge, even under unitary dynamics. For example, a fractional exponent $3/2$ has been reported in the early decay of the survival probability of sharply localized wavepackets undergoing free evolution \cite{SPK12}. 

Provided that it exists, as is generally the case, the short time asymptotics (\ref{stsurva}) is consistent with a Gaussian decay up to $\mathcal{O}(t^3)$,
\beqa
S(\beta, t)=\exp\left(-\frac{t^2}{\tau_Z^2}\right).
\eeqa
Yet, for systems strongly perturbed away from equilibrium, the validity of this Gaussian decay can extend well beyond the short-time regime \cite{Santos14}.

Using the definition of the specific heat in terms of the energy fluctuations in a thermal state, one finds the estimate in \cite{PR15}, that leads to the identification of the Zeno time 
\beqa
\tau_Z=\beta\sqrt{\frac{k_B}{c_V}}.
\eeqa

\subsection{Long-time nonexponential decay}
\label{sec.LongTimeDecay}

For  systems with a continuum spectrum, Fock and Krylov showed that the survival probability of an arbitrary initial state vanishes identically at $t\rightarrow \infty$ \cite{FK47}.
The theorem requires the existence of a ground state so that the Hamiltonian is bounded from below, and follows from the analytic properties of the survival probability and its Fourier transform, see Appendix \ref{AppendixFock}.

By contrast in systems with a discrete spectrum the quantum recurrence theorem (QRT) \cite{BL57,Schulman78} states that there exist approximate revivals, in the sense that according to  various notions of distance in Hilbert space  the time-dependent state and the initial state become arbitrarily close for some time of evolution.

As a result, we can anticipate a completely different long-time behavior depending on whether the spectrum is continuous or discrete.

In systems with discrete spectrum, the exact expression of the survival probability (\ref{survateq}) can we rewritten as
\beqa
\label{survatdec}
S(\beta,t)=\sum_{n}p_n^2+\sum_{n\neq m}p_np_m
\cos[(E_n-E_m)t].
\eeqa
where $p_n=\exp(-\beta E_n)/Z$ are the Boltzmann factors defining the occupation probability in the canonical thermal density matrix that results from tracing the degrees of freedom, e.g., of the right mode, $\rho_L=\tr_R[\psi(\beta,0\ra\la\psi(\beta,0)|]=\sum_np_n|n\ra\la n|$.
The constant term in the right-hand side of (\ref{survatdec}) is  given by the purity of the canonical state
\beqa
\mathcal{P}(\rho_L)=\tr \rho_L^2=\sum_{n}p_n^2,
\eeqa
and equals the inverse participation ratio, providing a measure of the delocalization of the thermal state over the spectrum of the Hamiltonian. The purity satisfies
\beqa
\frac{1}{d}\leq \mathcal{P}(\rho_L)\leq 1,
\eeqa
where $d={\rm dim}\mathcal{H}$ is the dimension of the Hilbert space of the left mode $\mathcal{H}$, the upper bound is only reached when  $\rho_L$ is a pure state (i.e. at zero temperature in the absence of degeneracy of the ground state) and the  lower bound follows from considering the maximally-mixed state associated with the infinite temperature limit, whose von Neumann entropy is $S[\rho_L(\beta\rightarrow\infty)]=\log d$. 
In a thermal state, the purity can be written in terms of the partition function as
\beqa\label{eq.Purity}
\mathcal{P}(\rho_L)=\sum_n\left(\frac{e^{-\beta E_n}}{Z(\beta)}\right)^2=\frac{Z(2\beta)}{Z(\beta)^2}.
\eeqa
It is the purity of $\rho_L$ that determines the long-time average of the survival probability,
\beqa\label{eq.PurityBound}
\overline{S}(\beta)=\lim_{T\rightarrow\infty}\frac{1}{T}\int_0^T S(\beta,t)=\mathcal{P}(\rho_L).
\eeqa

By contrast, the decay dynamics is manifestly different when the spectrum is continuous.
To study  the long-time asymptotic of the survival probability we introduce  the density of states 
\beqa
\rho(E)=\sum_nN_E\delta(E-E_n), 
\eeqa
using the  degeneracy $N_E$ and rewrite the survival amplitude as
\beqa
\label{survat}
A(\beta,t)=\frac{1}{Z}\int_{E_0}^\infty dE\rho(E)e^{-\beta E-i Et}.
\eeqa
Following  Khalfin \cite{Khalfin57,FGR78}, we note that $A(\beta,t)$ is then the Fourier transform of a function with support on the half energy axis, $\tilde{\rho}(E)\equiv\rho(E)\Theta(E-E_0)$. Taking the point of view of the inverse transform, in order for the spectral density to vanish for energies less than $E_0$, the Paley-Wiener theorem implies that the integral
\beqa
\int_{\mathbb{R}} dt\frac{|\log|A(\beta,t)||}{1+t^2}<\infty.
\eeqa
be finite. For this to be true it has to be the case that 
\beqa
S(\beta,t)\geq Ce^{-\gamma t^q},
\eeqa
with $C,\gamma>0$ and $q<1$. In other words the long-time asymptotics are characterized by a slower than exponential decay.

The specific form of the long-time decay is set by the low-energy behavior of the spectral density. 
Under quite general conditions, a power-law decay holds, as has been demonstrated in single-particle \cite{FGR78}, few-body \cite{TS11} and many-body systems \cite{delcampo11,delcampo16,Santos14, Kamenev16}. 
Indeed, if $\rho(E)\sim E^k$ near $E=E_0$, the survival probability at long times scales as
\beqa
\label{eqltpl}
S(\beta,t)\propto(t^2+\beta^2)^{-(k+1)}.
\eeqa

Physically, the origin of long-time deviations from exponential decay can be traced back to the possibility of state reconstruction under unitary quantum dynamics. To argue this, let us introduce the projector onto the subspace spanned by the TDS
\beqa
\hat{P}=|\psi(\beta)\ra\la\psi(\beta)|, \qquad \hat{P}^2=\hat{P},
\eeqa
as well as its orthogonal complement $\hat{Q}=1-\hat{P}$.
Following Ersak \cite{Ersak69,FGR78}, we insert the resolution of the identity $1=\hat{P}+\hat{Q}$ after a time of evolution $0\leq \tau\leq t$. The survival amplitude can then be exactly written as
\beqa\label{eq.ErsakEq}
A(\beta,t)=A(\beta,t-\tau)A(\beta,\tau)+M(\beta,t,\tau),
 \eeqa
where the last term represents a memory term 
  \beqa
 M(\beta,t,\tau)=\la \Psi(\beta,0)|U(t,\tau)\hat{Q}U(\tau,0)|\Psi(\beta,0)\ra,
 \eeqa
and $U(t,t')$ is the unitary time evolution operator. From Eq. (\ref{eq.ErsakEq}), we see that whenever $M(\beta,t,\tau)$ vanishes, $A(\beta,t)$ follows and exponential decay. Furthermore, we can analyze the contributions of the different terms appearing in this expression \cite{Muga06,delcampo16}. Defining $A_P(\beta,t)=A(\beta,t-\tau)A(\beta,\tau)$, the survival probability reads
\beqa
S(\beta,t)=|A_P|^2+|M|^2+2{\rm Re}[A_P^*M].
\eeqa
In systems with a continuous spectrum bounded from below, the long-time asymptotic behavior of $S(\beta,t)$ is governed by $|M|^2$. As a result, it is therefore associated with rare events in which the dynamics first leads to ``decay products'', identified with the component $\hat{Q}|\psi(\beta,\tau\ra$,  that subsequently evolve to reconstruct the initial state at time $t$. 
The suppression of the memory term, e.g. via nonunitary dynamics, leads to exponential behavior, as seen from the Ersak equation (\ref{eq.ErsakEq}).

\section{Examples}\label{sectionEx}

We next focus on several examples with both a discrete and continuous spectrum, to illustrate the salient features of quantum decay. We proceed in increasing order of complexity, starting with the simplest imaginable case, the harmonic oscillator. Interestingly, as we show, the same survival probability is shared by the conformal quantum mechanics of the $0+1$ CFT described by the $xp$ model\footnote{This happens despite the fact that the xp model saturates \cite{JaviUnpub} the chaos bound of \cite{Maldacena15}. This should be seen as an illustration of the fact that any model that shares the conformal symmetry and its breaking patters with that of the SYK model will exhibit maximal chaos in this sense \cite{Jensen:2016pah}.} in AdS$_2$. We next illustrate the case of an integrable Hamiltonian on the example of the Calogero-Sutherland model. Next we move on to the dynamics of the survival probability in random matrix theory, giving the exact finite-$N$ answer for the case of the Gaussian unitary ensemble. We end with a discussion of the survival amplitude in field theories with holographic duals.

\subsection{Quantum harmonic oscillator and the $xp$ model in AdS$_2$}\label{sec.SHOxp}
The one-dimensional harmonic oscillator
is a simple but illustrative example.
The spectrum $E_n=\om(n+1/2)=E_0+n\om $ is unbounded above ($n=0,1,2,\dots$) leading to
\beqa
\label{sho}
S(\beta,t)=\frac{\cosh(\om\beta)-1}{\cosh(\om\beta)-\cos(\om t)}.
\eeqa 
As a result, $S(\beta,t)$ is an oscillatory function of time with a single nominal frequency $\om$.
This is a rather special behavior specific to the TDS of the harmonic oscillator, 
due to the specific form of the probability amplitude in the $n$-th mode $c_n=\exp(-\beta E_n/2)/\sqrt{Z(\beta)}$.
The short-time asymptotics reveal the energy fluctuations
\beqa
S(\beta,t)=1-\frac{\om^2 t^2}{2[\cosh(\om\beta)-1]}+\mathcal{O}(t^3).
\eeqa
We also note that in the low temperature limit the first two terms of the expansion vanish and $S(\beta,t)=\mathcal{O}(\beta^2)$. In particular,
\beqa
S(\beta,t)=-\frac{\om^2 \beta^2}{2[\cosh(\om\beta)-1]}+\mathcal{O}(\beta^3).
\eeqa

We further point out that the state $|\psi(\beta,t)\ra$ is nothing but a two-mode squeezed state, that is generally defined by the action of the two-mode squeezing operator $S_{LR}(\zeta)$ on the vacuum state \cite{BR05}
\beqa
|\zeta_{LR}\ra&=&S_{LR}(\zeta)|0_L,0_R\ra,\nonumber\\
&=&\exp(-\zeta \hat{a}_L^\dag\hat{a}_R^\dag+\zeta^*\hat{a}_R\hat{a}_L)|0_L,0_R\ra,
\eeqa
with an arbitrary complex number $\zeta=r\exp(i\varphi)$.
In the Fock basis, the explicit representation reads
\beqa
|\zeta_{LR}\ra={\rm sech} r\sum_n\left(-e^{i\varphi}{\rm tanh} r\right)^n|n_L,n_R\ra.
\eeqa
We can rewrite the time-dependent state  
\beqa
|\psi(\beta,t)\ra=\exp(-iE_0t)|\zeta_{LR}(t)\ra
\eeqa
with the identification of $\zeta(t)$ via
\beqa
{\rm tanh} r&=&\exp(-\beta \om/2),
\varphi=\om t.
\eeqa
The survival probability can then alternatively be expressed as 
\beqa
S(\beta,t)=|\la\zeta_{LR}(0)|\zeta_{LR}(t)\ra|^2,
\eeqa
which explains the appearance of a single frequency in the explicit expression (\ref{sho}). Clearly, $S(\beta,t)$ is a periodic function of time, $S(\beta,t)=S(\beta,t+2\pi/\om)$.

As it turns out, the survival probability of the harmonic oscillator is shared by the $xp$ model in AdS$_2$ \cite{molina_sierra13}. 
The corresponding Hamiltonian  reads
\beqa
\label{xp1}
H = w(x) (p + \ell_p^2/p),
\eeqa
where $x$ and $p$ are the position and momentum of a particle moving in the real line, $\ell_p$ is a parameter with the dimension of momenta, and $w(x)$ is an arbitrary positive function.  The  $xp$ model describes the motion of a relativistic  particle moving in a 1+1 spacetime whose  metric is determined by $w(x)$. While the Riemann scalar curvature  vanishes identically for the linear  potential $w(x)=x$, for $w(x) = w_0  \cosh(x/R)$ (with $w_0$ and $R$ being positive constants) yields a spacetime with constant negative curvature ${\cal R} = - 2/R^2$, which corresponds to an anti-de-Sitter spacetime  (AdS$_2$), with  radius $R$ \cite{molina_sierra13}. 
The classical Hamiltonian (\ref{xp1}) can be quantized in terms of the  following normal ordered operator 

\beqa
\label{xp2}
\hat{H} = \sqrt{w(x)}  \,    \hat{p} \,   \sqrt{w(x)}   +   \ell_p^2  \sqrt{w(x) }   \hat{p}^{-1}   \sqrt{w(x)},
\eeqa
where $\hat{p} = - i \hbar d/dx$ and $\hat{p}^{-1}$ is the pseudo-differential operator that acts of wave functions as $(  \hat{p}^{-1}  \psi)(x) = - \frac{i}{\hbar}  \int_x^\infty dy \, \psi(y)$.

The spectrum of  $\hat{H}$ has positive and negative eigenvalues, whose absolute values are given by a  harmonic oscillator spectrum,
\beqa
\label{xp3}
E_n = \hbar \omega \left(n + \kappa +\frac{1}{2}\right), \quad \kappa =\frac{R\, \ell_p}{\hbar}, \quad \omega = \frac{2\, w_0}{R}.
\eeqa

In \cite{molina_sierra13} it was shown how the symmetry group $SO(2,1)$ can be realized in the $xp$-AdS$_2$ model by building the generators of this group, both in the classical as in the quantum theory.  Namely, the $xp$ model realizes two  infinite dimensional representations of $SO(2,1)$ corresponding to the positive and  negative branches of the spectrum which are related by complex conjugation. The group $SO(2,1)$ describes  the symmetry of conformal quantum mechanics (CFT$_1$), as introduced by \cite{J72},  and analyzed in great detail in \cite{AFF}.

The survival probability in the CFT$_1$ realized by the $xp$ model then reads as
\beqa
\label{eq:surv_xp1}
S(\beta,t)=\left|\frac{Z_{CFT_1}(\beta,t)}{Z_{CFT_1}(\beta)}\right|^2 = \frac{\cosh ( \beta\omega) - 1}{\cosh ( \beta\omega) - \cos (\omega t)}, 
\eeqa
i.e., it is precisely given by (\ref{sho}).

\subsection{Rational Calogero-Sutherland model}\label{eq.CSmodel}
The Calogero-Sutherland model (CSM) describes one-dimensional particles subject to inverse-square interactions. 
In the presence of a harmonic potential \cite{Calogero71,Sutherland71}, the Hamiltonian reads
\begin{eqnarray}
 H=\sum_{i=1}^{N}\left[-\frac{1}{2}\frac{\partial^2}{\partial z_i^2}+\frac{1}{2}\omega^2 z_{i}^{2}\right]+\sum_{i<j}\frac{\lambda(\lambda-1)}{(z_i-z_j)^2}\ ,
 \label{hcsm}
\end{eqnarray}
where $\lambda$ is the coupling constant. The CSM is then equivalent to an ideal gas obeying fractional exclusion statistics and includes noninteracting bosons and fermions as well as more general particles known as  geons or Haldane anyons \cite{Haldane91,Wu94}. The rational Calogero model of $N$ particles shares the spectrum
with the two-dimensional $SU(N)$ super Yang-Mills on a cylinder and has been used to analyze the micro states of four-dimensional eternal blackholes \cite{Claus98,GT99,LN16}.

Surprisingly, the only effect of the coupling constant $\lambda$ in the spectrum is to renormalize the zero-point energy 
\beqa
E_0=\frac{\om}{2}N[1+\lambda(N-1)].
\eeqa

The survival probability of this system has been studied under  arbitrary modulations in time of the frequency $\om$ \cite{delcampo16}.
Here, we are interested in its value for the TDS that can be directly obtained by analytic continuation of the partition function. Murthy and Shankar showed a duality relating the partition function of the CSM to that of noninteracting bosons and fermions \cite{MS94}. Using it, it is possible to derive the explicit partition function in the canonical ensemble \cite{JBD16}, and ultimately, the survival probability,
\beqa
S(\beta,t)=\prod_{k=1}^N\frac{\cosh( k \om\beta)-1}{\cosh( k\om\beta)-\cos(k \om t)}.
\eeqa
The result is still surprisingly simple when compared with the survival probability of an arbitrary coherent superposition of many-body eigenstates. First, $S(\beta,t)$ is independent of the coupling constant whose only effect in the spectrum is the renormalization of zero-point-energy.  While the survival probability of the TDS is independent of $E_0$ and $\lambda$, this is no longer the case under more complex scenarios, e.g., involving quench dynamics \cite{delcampo16}. For $N=1$ one naturally recovers the behavior of a single-harmonic oscillator. For $N>1$, 
$S(\beta,t)$ acquires a more complex modulation with frequencies $k\om$ ($k=1,\dots,N$). The survival probability is generic for an ideal gas of bosons, fermions or geons. Indeed, it brings out the fact that the CSM spectrum is equivalent to that of $N$ renormalized harmonic oscillators, up to the zero-point energy \cite{Calogero71,Kawakami93,GP99}. 
As in the case of the single-harmonic oscillator, the survival probability is a periodic function of time, with period
$2\pi N/\om$.

\subsection{Random Matrix Hamiltonians}\label{RMTH}
\subsubsection{Survival probability at large $N$}
% ---------------- FIG. 1 begin ----------------
%
\begin{figure*}[t]
\begin{center}
\includegraphics[width=0.33\linewidth]{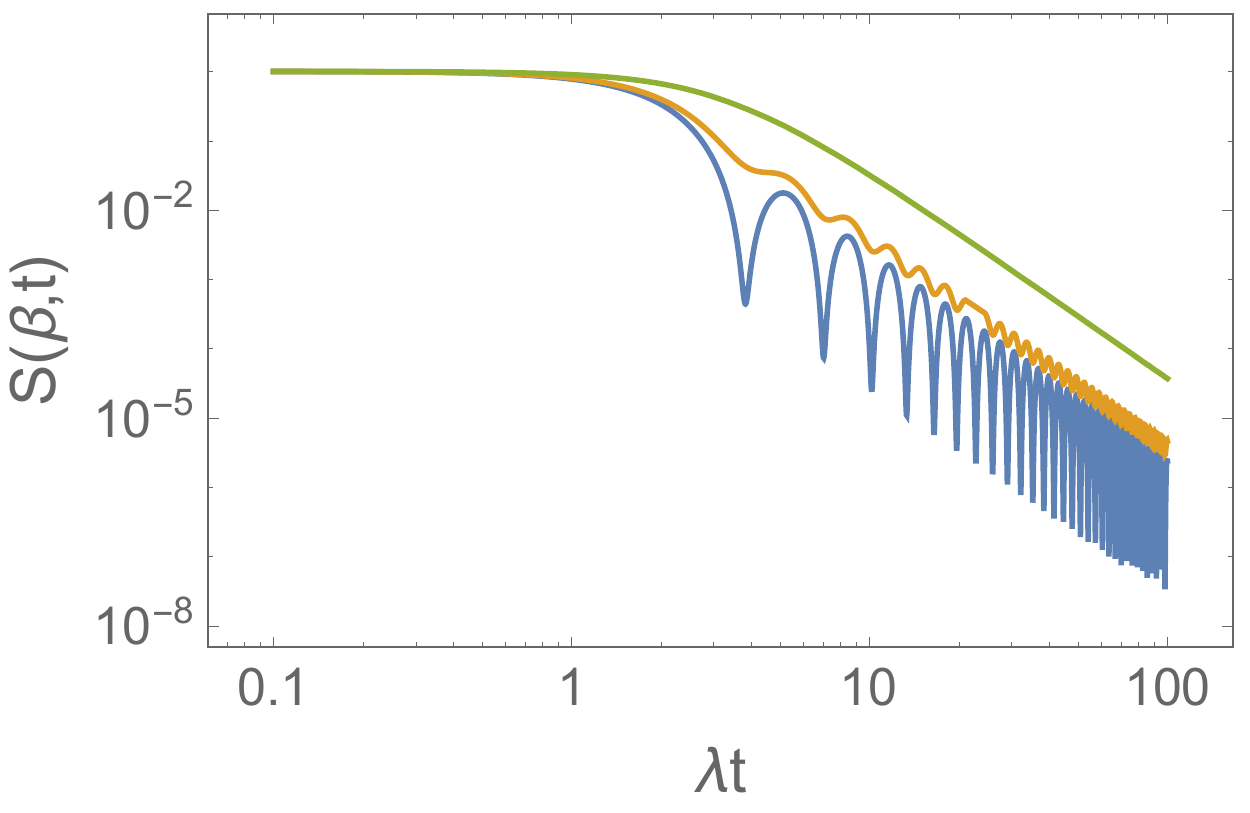}
\includegraphics[width=0.315\linewidth]{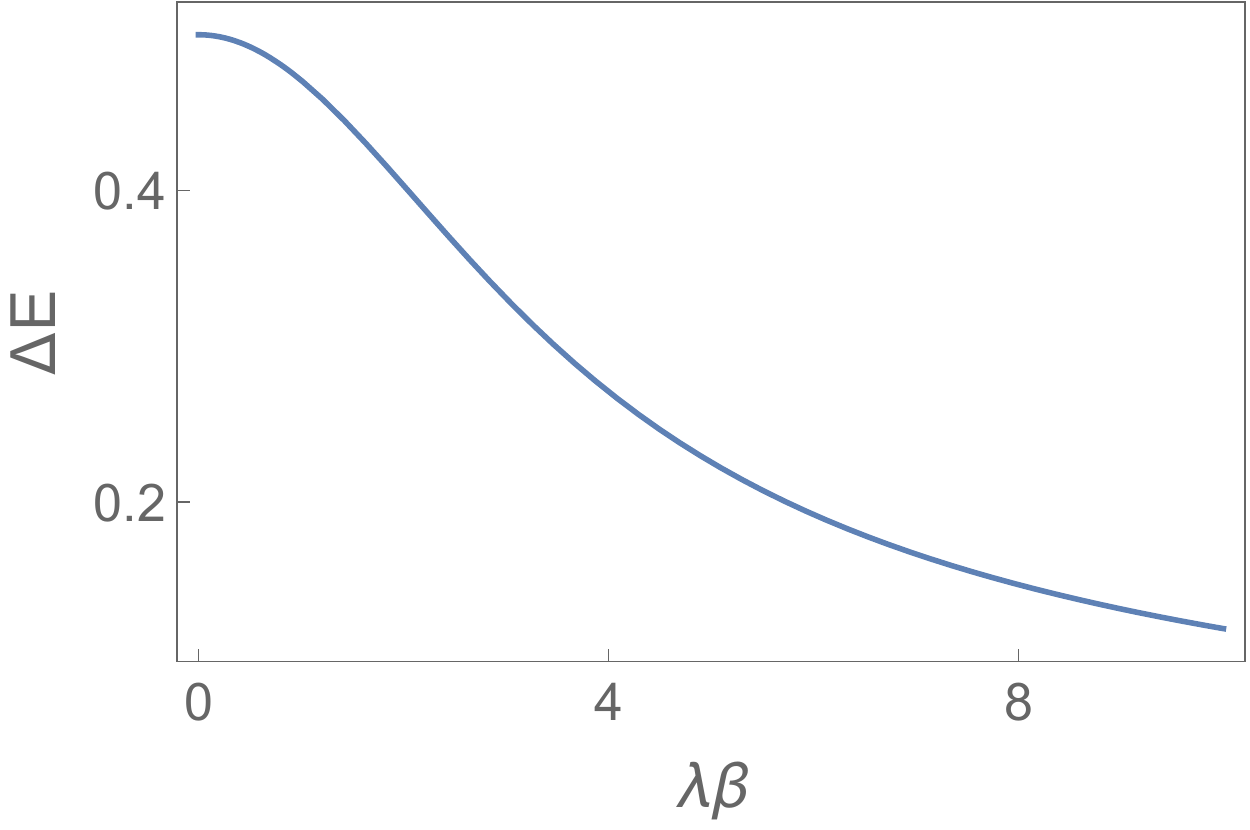}
\includegraphics[width=0.32\linewidth]{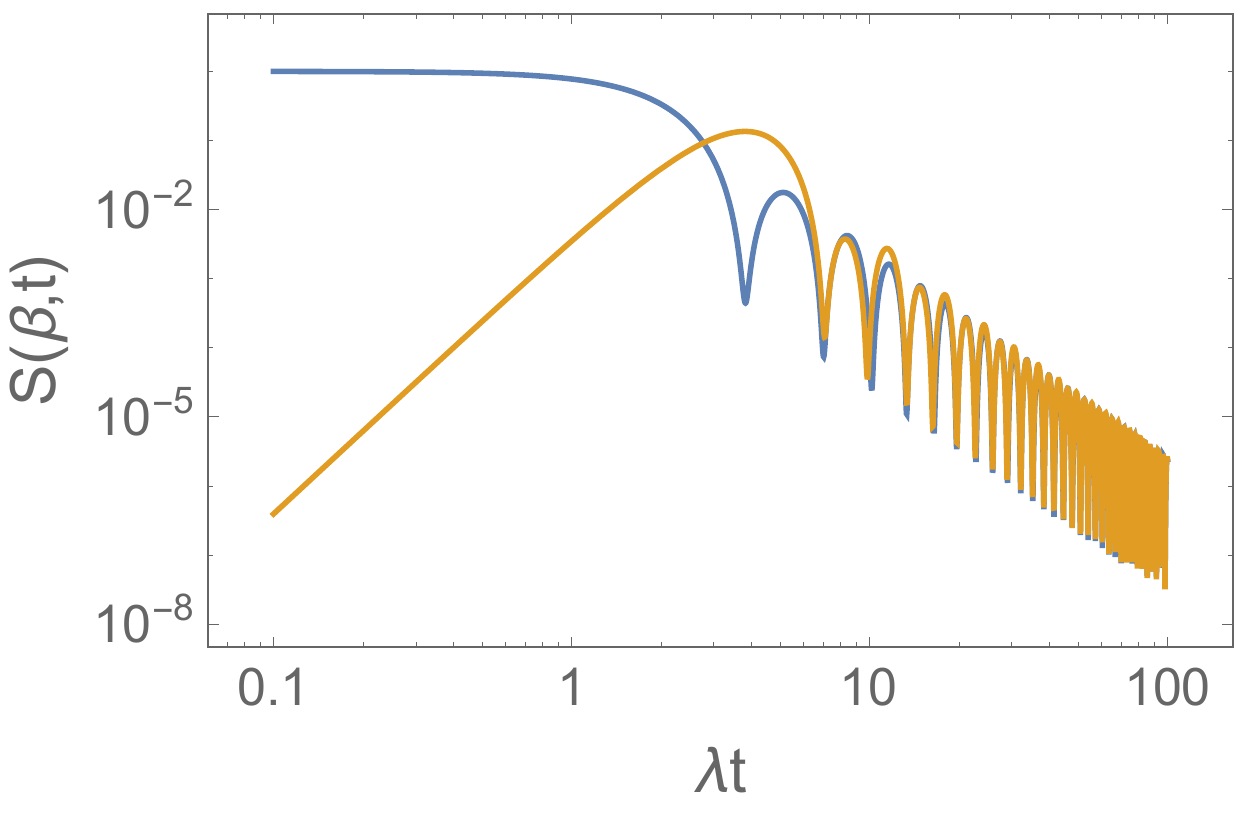}
\end{center}
\caption{\label{survatWignerFig}  {\bf Survival probability of the thermofield double state in the GUE.} (a) The exact survival probability of the TDS is displayed as a function of time for $\lambda\beta=0.1,1,3$, from bottom to top.  During its decay, $S(\beta,t)$ exhibits oscillations as a function of time, the amplitude of which diminishes with the temperature. In the low temperature regime the decay is approximately  monotonic, and governed by a power law $t^{-3}$.
(b) Energy fluctuations in a TDS as a function of the inverse temperature $\beta$.
(c) Long-time quantum reconstruction of the TDS under unitary dynamics. The exact survival probability of the TDS is compared in logarithmic scale with the probability of the memory term. At long-times of evolution the memory term accurately reproduces the survival probability indicating that the exact evolution is consistent with the decay of the TDS in a classical probabilistic sense at intermediate times and its subsequent reconstruction. The dynamics is associated with a GUE random matrix ensemble at large $N$ with a density of states described by Wigner semicircle law with $\lambda\beta=0.1$. The memory term is evaluated at $\tau=t/2$.
}
\end{figure*}

We now turn our attention to the survival probability of a TDS when the dynamics is generated by a full random matrix Hamiltonian. In particular, we consider Hamiltonians that are described by the Gaussian Unitary Ensemble (GUE), i.e., the Dyson $\beta$-ensemble with $\beta=2$.

In the asymptotic limit  for matrices of large rank, the density of states obeys Wigner's semicircle law \cite{MehtaBook}
\beqa
\rho(E)&=&\lim_{N\rightarrow\infty}\left\la\sum_{n=1}^N\delta(E-E_n)\right\ra_{\rm GUE},\nonumber\\
&=&\frac{\lambda}{\pi}\sqrt{1-\left(\frac{E}{\lambda}\right)^2},
\eeqa
in the interval $E\in[-\lambda,\lambda]$ and $\rho(E)=0$ elsewhere.
We choose the (dimensionless) Wigner radius  $\lambda=\sqrt{2N}$ so that the normalization condition reads
\beqa
\int_{-\lambda}^\lambda\rho(E)dE=\frac{\lambda^2}{2}=N.
\eeqa
Explicit evaluation via the integral representation of the survival amplitude (\ref{survat})  leads to
 \beqa
A(\beta,t)&=&\frac{\lambda J_1[\lambda(t-i\beta)]}{Z(t-i\beta)},\\
S(\beta,t)&=&\frac{\lambda^2 J_1[\lambda(t+i\beta)]J_1[\lambda(t-i\beta)]}{Z^2(t^2+\beta^2)},\eeqa
where $J_1(x)$ is the Bessel function of first kind of first order, and the partition function reads
\beqa
Z=Z(\beta,0)=\frac{\lambda I_1(2\beta)}{\beta},
\eeqa
with $I_n(x)$ denoting the modified  Bessel function of first kind and order $n$.
 The infinite-temperature limit has recently been reported in \cite{Cotler16},  
 and a closely related analysis was discussed in \cite{Santos14}, in the study of thermalization of isolated many-body quantum systems.

The short-time asymptotic behavior of $S(\beta,t)$ takes the form of (\ref{stsurva}) with the identification of the Zeno time
\beqa
\label{enflucsc}
\frac{\lambda^2}{\tau_Z^2}=
1-\frac{3I_2(\lambda\beta)}{\lambda \beta I_1(\lambda\beta)}
-\frac{I_2(\lambda\beta)^2}{ I_1(\lambda\beta)^2}\le \frac{1}{4},
\eeqa
which is the inverse of the energy variance, showing that $\Delta E\leq\lambda/2$, with the equality holding at infinite temperature.
The long-time asymptotic simply reads 
\beqa
S(\beta,t)=\frac{\lambda}{\pi Z^2}\frac{\cosh(2\lambda\beta)-\sin(2\lambda t)}{t^3}+\mathcal{O}(t^{-4}).
\eeqa

The full quantum decay of the survival probability as a function of time is shown in Fig. \ref{survatWignerFig}(a). 
In an early stage the decay is parabolic in time and characterized by the energy fluctuations whose inverse set the Zeno time, see Fig. \ref{survatWignerFig}(b). As the time of evolution goes by,  a power-law behavior sets in and $S(\beta,t)\propto t^{-3}$. 
This behavior can be traced back to the compact support of the  Wigner's semicircle law in the light of Paley-Wiener theorem. 
Actually, given that near the low-energy edge $E=-\lambda$, the density of states scales as $\rho(E)\sim (E+\lambda)^{1/2}$, 
the power-law $t^{-3}$ follows from Eq. (\ref{eqltpl}).
Yet, at high temperatures, the TDS in Eq. (\ref{eq.TFD}) involves a coherent superposition over many modes and during its decay, the survival probability   exhibits an oscillatory behavior superimposed on the power law. As the temperature is decreased, the amplitude of the oscillation diminishes and the decay is eventually monotonic. 
Figure  \ref{survatWignerFig}(c)
shows that the long-time behavior at arbitrary temperature is dominated by the memory term and hence it is associated with the reconstruction of the TDS from the decay products. The survival probability is then dominated by an evolution during which the initial TDS is fully scrambled at an intermediate time $\tau$ in the sense that  the time evolving state is completely orthogonal to the initial state, 
\beqa
\hat{P}|\psi(\beta,\tau)\ra=0, \hat{Q}|\psi(\beta,\tau)\ra=|\psi(\beta,\tau)\ra,
\eeqa
yet the decay products $ \hat{Q}|\psi(\beta,\tau)\ra$ subsequently reconstruct the initial state at the later time $t$.

\subsubsection{Exact fidelity at finite $N$}\label{sec.ExaFi}
In fact we can go much further and analytically evaluate the fidelity decay in the RMT for any $N$ using the method of orthogonal polynomials.  For previous work on the spectral form factor at finite $N$, see \cite{brezin1997spectral}. We want to compute the exact survival amplitude of a TDS  in the GUE, in other words, the spectral form factor
% better this way?
\beqa\label{eq.GUEZZs}
g(\beta,t)=\langle Z(\nu) Z(\nu^*)\rangle &=& \langle Z(2\beta)\rangle + \left\langle\sum_{i\neq j}e^{-\beta(E_i + E_j) - i t(E_i-E_j)}\right\rangle, \nonumber\\
&=&\langle Z(2\beta) \rangle +  \int dE_1 dE_2 \left\langle \rho^{(2)} (E_1,E_2)\right\rangle e^{-\nu E_1 - \nu^* E_2}\nonumber,
\eeqa
where the last equality defines the GUE averaged two-level correlation function $\rho^{(2)} (E_1,E_2)$. 
In order to deal with the integral over the two-level correlation function in (\ref{eq.GUEZZs}), we express its average $\langle\rho^{(2)}(E_1,E_2)\rangle$  in terms of the connected two-level correlation function $\rho_c^{(2)}(E_1,E_2)$ as
\beq
\left\langle\rho_c^{(2)}(E_1,E_2) \right\rangle=  \left\langle\rho^{(2)}(E_1,E_2) \right\rangle - \langle\rho(E_1)\rangle\langle\rho(E_2)\rangle\,.
\eeq
As a result, we identify three terms in the spectral form factor
\beqa
\label{g3}
 g(\beta,t)=\langle Z(2\beta) \rangle +|\langle Z(\beta,t)\rangle|^2+g_c(\beta, t),
\eeqa
where $g_c(\beta, t)$ is the double complex Fourier transform of the connected contribution
\beq\label{eq.TwoLevelConnectedCorrelation}
g_c(\beta, t) = \int dE_1 dE_2 \left\langle \rho_c^{(2)} (E_1,E_2)\right\rangle e^{-\nu E_1 - \nu^* E_2}.
\eeq
Let us now compute  each of the terms appearing in this expression.
We begin by computing the exact analytically continued partition function
\beq
Z(\beta, t)=\sum_n e^{-(\beta +it ) E_n}\nonumber
\eeq
which we rewrite as the integral
\beqa
Z(\beta, t)=\int_{\mathbb{R}} dEe^{-\nu E}\sum_{n=1}^N\delta(E-E_n)\,.
\eeqa
where the quantity $\nu$ appearing in the exponent is the analytically continued inverse temperature, $\nu = \beta + it$. This expression still needs to be averaged over the random matrix ensemble. In fact, it is known \cite{MehtaBook} that the exact eigenvalue density averaged over the GUE is given by
\beqa
\label{rhoGUE1}
\langle\rho(E)\rangle=\sum_{j=0}^{N-1}\varphi_j(E)^2,
\eeqa
where we introduced the harmonic oscillator eigenfunctions
\beqa
\varphi_j(E)=\frac{1}{(2^jj!\sqrt{\pi})^{1/2}}e^{-\frac{E^2}{2}}H_j(E).
\eeqa
These play the role of orthonormal polynomials for the unitary matrix ensemble, in terms of which we will be able to perform all of the calculations in this section \footnote{We have adopted the normalization of  \cite{MehtaBook}, where the density of states approaches the limiting Wigner distribution as $\rho_N(E) \rightarrow \frac{1}{\pi}\sqrt{2N - E^2}$.}.  
After some algebra (see Appendix \ref{sec.ExactAmplitude}), we arrive at the compact expression
\beqa\label{eq.ExactGUEfidelityDecay}
\langle Z(\beta,t)\rangle = e^{\frac{\nu^2}{4}}L_{N-1}^1(-\tfrac{\nu^2}{2})\,,
\eeqa
that allows us to compute the first two terms in  (\ref{g3}).
We note the similarity with the expectation value of a circular (BPS) Wilson loop in the ${\cal N}=4$ SYM theory. In fact this is no coincidence, as we explain in appendix \ref{sec.MapToWilson}. By contrast to Wigner's semicircle law, the density of states $\langle\rho(E)\rangle$ (\ref{rhoGUE1}) no longer has a compact support on the energy axis and takes non-zero values in the full real line. As a result, the decay of $\langle Z(\beta,t)\rangle$ is not restricted to be subexponential and indeed, following a power-law regime with an envelope $1/t^3$, this term is highly suppressed as its long-time asymptotic behavior reads
\beqa
|\langle Z(\beta,t)\rangle|^2\sim\frac{4^{N-1}}{\Gamma(N)^2}e^{-\frac{t^2-\beta^2}{2}}t^{4(N-1)}.
\eeqa
To compute the third term in (\ref{g3}), we note that $\left\langle \rho_c^{(2)} (E_1,E_2)\right\rangle$  can be expressed conveniently in terms of normalized Hermite polynomials \cite{MehtaBook},
\beq\label{eq.TwoLevelCorrelationHermite}
\left\langle\rho_c^{(2)}(E_1,E_2) \right\rangle= -\left( \sum_{j=0}^{N-1} \varphi_j(E_1) \varphi_j(E_2)\right)^2\,.
\eeq
We then find that the double complex Fourier transform of the connected contribution
\beq\label{eq.TwoLevelConnectedCorrelation}
g_c(\beta, t) = \int dE_1 dE_2 \left\langle \rho_c^{(2)} (E_1,E_2)\right\rangle e^{-\nu E_1 - \nu^* E_2}
\eeq
can again be evaluated exactly (see Appendix \ref{sec.ExactAmplitude}). 
The result can be expressed compactly,
\beq\label{eq.connectedFormFactorGUE}
g_c(\beta, t) = e^{\frac{1}{4} \left(\nu^2 + \bar\nu^2  \right)}\sum_{j,k=0}^{N-1}\left( \frac{\nu}{\bar\nu} \right)^{k-j} \left|\psi_{jk}\left(-\frac{\nu^2}{2}\right)\right|^2\,,
\eeq
where the function $\psi_{jk}(x)$ is defined in terms of the confluent hypergometric function ${}_1F_1(a;b;x)$,
\beqa
\psi_{jk}(x)&=&\frac{{}_1F_1\left(-j; 1 + k - j; x\right)}{ \Gamma\left(1 + k -j \right)}=\frac{\Gamma(j+1)}{\Gamma(k+1)}L_{j}^{k-j}(x).
\eeqa
Let us remark that this exact expression can be mapped to the expectation value of two Wilson loops in the Gaussian matrix model, as shown in appendix (\ref{sec.MapToWilson}).

\begin{figure*}[t]
\begin{center}
a)\includegraphics[width=0.29\linewidth]{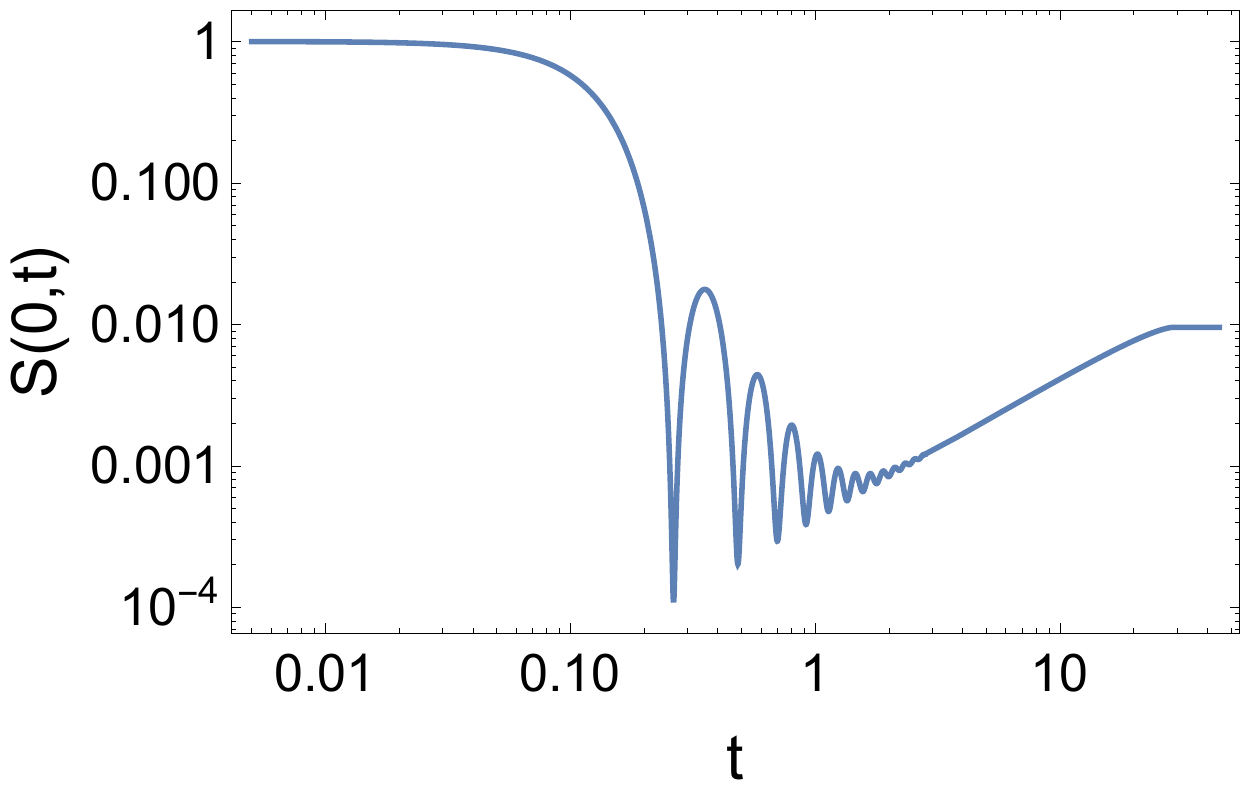}\hskip1em b)\includegraphics[width=0.29\linewidth]{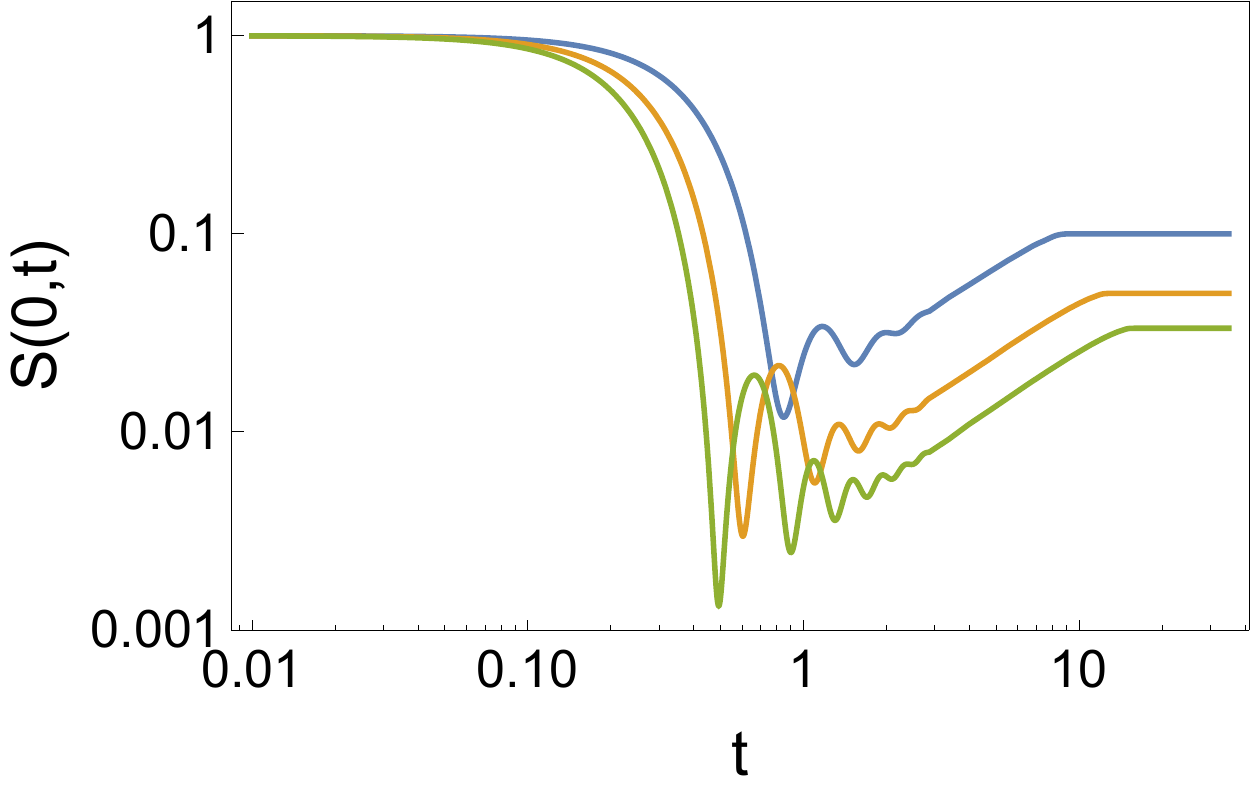}\hskip1em c)\includegraphics[width=0.29\linewidth]{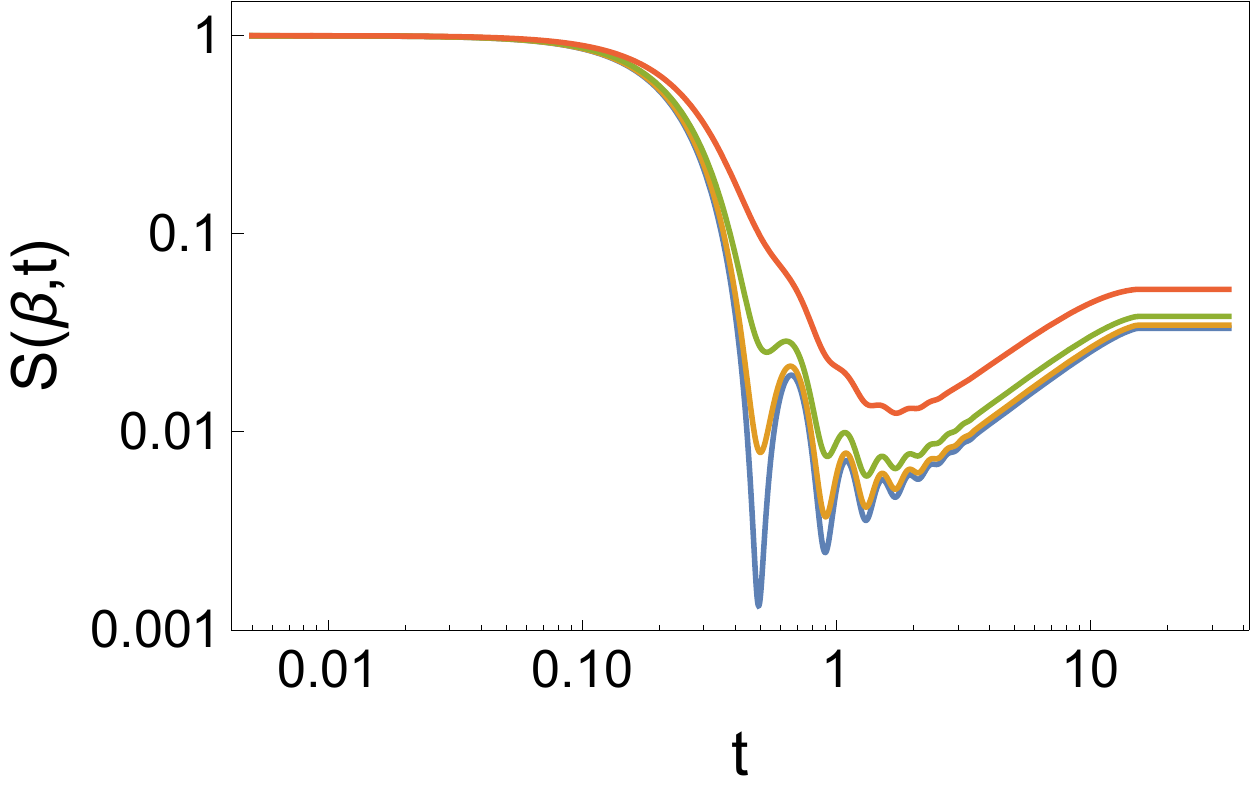}
\end{center}
\caption{\label{fig.ZZsAnalytic}  {\bf Exact survival probability in the GUE}. a) Zero-temperature for GUE of size $N=105$: A period of decay and approximate revivals with a power-law envelope that scales as $1/t^3$ terminates at  a dip, followed by a linear rise that saturates at a plateau of order $1/N$. b) Comparison of zero-temperature curves for $N=10$ (blue), $N=20$ (orange) and  $N=30$ (green). c) Comparison of different curves with $N=35$ and temperature $\beta = 0,0.05,0.1,0.2,0.4$, from bottom to top.
 As the inverse temperature increases, the initial decay approaches a straight line, while the dip, ramp and plateau phase remain relatively stable.}
\end{figure*}

We now have all the information needed to reconstitute the exact spectral form factor, or equivalently the survival probability, by adding the disconnected component as well as the $Z(2\beta)$ contribution appearing in (\ref{eq.GUEZZs}) to the connected form factor (\ref{eq.connectedFormFactorGUE}). The resulting behavior\footnote{Here we use the annealed GUE averaged decay amplitude, that is $\langle S(\beta,t)\rangle =\langle g(\beta,t)\rangle/\langle Z(\beta)\rangle^2 $.} of $S(\beta, t)$ is illustrated in Figure \ref{fig.ZZsAnalytic}. As required the resulting curve shows the characteristic dip, ramp and plateau behavior with a plateau height
\beq\label{eq.plateauHeight}
h_P \sim 1/Z(\beta) \sim 1/N\,.
\eeq

 Let us now analyse the result for the finite-$N$ spectral form factor in more detail. Firstly, let us point out that, for any finite $N$, the spectral form factor (\ref{eq.GUEZZs}) approaches the purity (\ref{eq.Purity}) at late times. This is a consequence of the fact that, as shown in Appendix \ref{sec.ExactAmplitude}$, \rho^{(2)}(E,E')$ can be expressed as a finite sum over (products of) integrals of the type (\ref{eq.MasterIntegral2}), which we repeat here for convenience:
 $${\cal I}_{mn}(\beta,t) = \int_{\mathbb{R}} e^{- it E - \beta E}\varphi_{m}(E)\varphi_n(E)\,.$$
These integrals involve the Fourier transform of an $L^1$-function, and thus tend to zero as $t\rightarrow \infty$, by the Riemann-Lebesgue Lemma (see Appendix for more detail). As a consequence the spectral form factor tends to the purity at late times, as claimed above. In fact this result extends to the limit $N\rightarrow \infty$, where $\rho^{(2)}(E,E')$ approaches the sine kernel \cite{MehtaBook}
$$\rho^{(2)}(E,E') \sim \left[\frac{\sin\left(2N(E-E') \right)}{2N(E-E')}\right]^2\,,$$ 
which again satisfies the conditions of the Riemann--Lebesgue Lemma. Notice that in this limit the purity, i.e.  the plateau height (\ref{eq.plateauHeight}) also goes to zero, so that the spectral form factor decays strictly to zero at late times.

From the exact expression we can discern three contributions, each of which is responsible for one of the characteristic shapes seen in Figure \ref{fig.ZZsAnalytic}. 
\begin{enumerate}
\item {\bf Gaussian decay}: Initially $g(t,\beta)$ is dominated by the disconnected contribution, so that the initial decay is governed by the behavior of the Laguerre polynomial. We find that the short-time asymptotics takes the form (\ref{stsurva}) with the Zeno time
\beqa
\frac{1}{\tau_Z^2}=\frac{1}{2}+\frac{\beta^2L_{N-3}^3(-\frac{\beta^2}{2})+L_{N-2}^2(-\frac{\beta^2}{2})}{L_{N-1}^1(-\frac{\beta^2}{2})}-\frac{\beta^2L_{N-2}^2(-\frac{\beta^2}{2})^2}{L_{N-1}^1(-\frac{\beta^2}{2})^2}.\nonumber
\eeqa
In contrast with the large $N$ expression (\ref{enflucsc}), energy fluctuations $\Delta E^2=1/\tau_Z^2$ do not vanish in the low temperature limit, and saturate at a constant value $\Delta E^2=1/2$ independent of $N$.
\item  {\bf Power law \& dip}: The initial Gaussian behavior gives way to a power-law decay, or at higher temperature to a set of oscillations with power law envelope
\beqa
S(\beta,t) \sim 1/t^3\,,
\eeqa
which eventually joins the increasing contribution coming from the connected piece $g_c(\beta, t)$ at the dip time $t_d \sim {\cal O}(1)$. The estimate for $t_d$ follows from equating the power-law decay $1/t^3$ coming from the disconnected part to the approximately linear rise $\sim t$ coming from the connected part.
\item {\bf Ramp and plateau}: The connected contribution to the spectral form factor $g_c(\beta, t)$ approaches zero from below and by itself decays completely at late times. By adding the purity this leads to an approximately linear rise at late times until the plateau value of $Z(2\beta)$ is approached.
\end{enumerate}
{\bf Remark on $N$ scaling of dip time}: In \cite{Cotler16} the dip time at large $N$ was estimated to scale like $\tilde{t}_d\sim \sqrt{N}$, while in our normalization the dip time is always ${\cal O}(1)$. We now explain the relationship between the two estimates. If we want to normalize our density of states in such a way that, as $N\rightarrow\infty$ it approaches the Wigner semi-circle supported on a compact interval $[-\lambda,\lambda]$ with $\lambda = {\cal O}(1)$ (for definiteness, let us take $\lambda = \sqrt{2}$), we find that we need to rescale energy as $\sqrt{N}\tilde E = E$ and correspondingly $t = \tilde t/\sqrt{N}$ and similarly for the temperature. The large $N$ limit is then taken as $N\rightarrow\infty$ keeping $\tilde E$ and $\tilde\beta, \tilde t$ fixed. This leads to the dip time estimate 
$$ N^{3/2}\tilde t_d^{-3}  \sim\tilde t_d /\sqrt{N} \qquad\longrightarrow \qquad \tilde t_d \sim \sqrt{N}$$
as in  \cite{Cotler16}. The rescaled variables are convenient in the infinite-$N$ analysis, effectively keeping typical energies $\tilde E\sim {\cal O}(1)$, but at finite $N$ our normalization keeps notation as uncluttered as possible.

\section{Survival probability in AdS$_{d+1}$/CFT$_d$}\label{sectionAdS}

A further interesting arena where we can say a great deal about the fidelity decay of the thermofield double state is holographic duality, or AdS/CFT.  As we shall see, the behavior of $g(\beta, t)$ in random matrix theory is actually qualitatively very similar to the holographic case. In the latter context the thermofield double state is dual to the eternal black hole \cite{Maldacena:2001kr} and the decay properties of $S(\beta,t)$ can be interpreted as a diagnostic of information loss  and recovery \cite{Cotler16,Dyer16}. In this context it is easiest to think of $S(\beta,t)$ in terms of the analytically continued partition function (\ref{eq.SpecFormFac}), which is therefore the partition function of the eternal black hole. %The spectral form factor -- or equivalently the fidelity decay -- has been considered \cite{Cotler16} in the SYK model dual to an AdS$_2$ theory of gravity. Furthermore \cite{Dyer16} have given a detailed analysis of the two-dimensional case, i.e. the partition function of and AdS$_3$ black hole. 

As in RMT the time evolution of the spectral form factor in holography is characterized by three phases. First we have a period of gaussian decay, followed by a power law (or a power-law envelope), both roughly speaking dominated by the dual semi-classical black hole including one-loop corrections in the bulk.
 At very late times this decay is believed to give way to behavior typically associated with RMT: on average we see a linear ramp up to a plateau at the latest times which persists until the latest times. Poincar\'e recurrences are not expected until times of order $t\sim e^{e^{\rm S}}$ \cite{BR03,BR04}. In summary, then, the analytic RMT curves in Fig. \ref{fig.ZZsAnalytic} give a surprisingly good description of the holographic spectral form factor as well. For the sake of completeness, let us give a few m ore details on the AdS/CFT case as well, before discussing unitarity bounds in that context.

\subsubsection{Fidelity decay in AdS$_{d+1}$/CFT$_d$}
In holography the leading result for the partition function, valid at large central charge, which we denote generically by $c$, is given by an appropriately regularized version of the on-shell action of the Euclidean black hole or whichever other classical saddle dominates in the bulk,  \cite{Hawking:1982dh}
 \begin{equation}
 Z(\beta) \approx e^{-S_E^{\rm saddle}(\beta)}\,.
\end{equation}
where the $\approx$ sign indicates the presence of the fluctuation determinant around the saddle which we take into account below.
Let us illustrate this with the case of AdS$_3$ where the leading saddle point is given by the so-called BTZ black hole, and reads
\begin{equation}
Z(\beta) = e^{\frac{\pi^2 c}{3\beta}},
\end{equation}
where $c = \frac{3\ell}{2G_N}$ is the central charge of the dual CFT$_2$, related to $\ell$, the radius of AdS$_3$, in units of the Planck length as stated. The survival probability in this approximation is therefore obtained simply by analytic continuation
\begin{equation}
S(\beta,t)=e^{-\frac{2\pi^2 c }{3\beta}\frac{t^2}{\beta^2+t^2}} \sim 1 - \frac{2\pi^2 c}{3\beta^3}t^2\,,\qquad (t\ll \beta)\,.
\end{equation}
We observe, therefore, that the holographic fidelity shows the non-exponential initial decay (\ref{stsurva}) with Zeno time
\begin{equation}\label{eq.HoloZenoTime}
\tau_Z^2 = \frac{3\beta^3}{2\pi^2 c} := \frac{1}{(\Delta E)^2}\,,
\end{equation}
which is suppressed in $1/c$ for large central charge.
Considering in addition the one-loop fluctuation determinant around the saddle \cite{Giombi:2008vd,Dyer16} gives the prefactor
\begin{equation}\label{eq.semiClassicalg}
S(\beta,t) \sim \frac{1}{t^6}e^{-\frac{2\pi^2 c}{3\beta}} \qquad \qquad({\rm large}\,\,\,t)\,,
\end{equation}
leading to algebraic decay (at late times). A better estimate of the decay of the spectral form factor in CFT$_2$ is obtained by including the images under $SL(2,\mathbb{Z})$ of the vacuum character, which results in a $1/t^3$ power law \cite{Dyer16}.

This semiclassical analysis is only sensitive to a coarse-grained density of states where all discreteness has been smoothed out. The resulting density of states again satisfies the conditions of the Riemann-Lebesgue lemma (see Appendix \ref{sec.CompleteDecay}) and accordingly $S(\beta,t)$ decays to zero. For a system with a finite Hilbert space this state of affairs would imply that information has been lost,  since the discreteness of the spectrum forbids such complete decay, as explained in Section \ref{sec.LongTimeDecay}. The timescales of this can be estimated as follows \cite{Dyer16} (see also  \cite{Cotler16} for a conjecture on the 4D SYM case):
one writes the exact quantum partition function of the dual CFT$_2$ as an expansion in analytically known Virasoro characters. Then, by {\it assuming} that the latest time behavior is dominated by random matrix theory, one infers -- on average -- a linear ramp up to an eventual plateau, as we have shown analytically in previous sections. By combining the detailed decay with the very late RMT behavior, one may estimate the `dip' time in a CFT$_2$ dual to the AdS$_3$ black hole as \cite{Dyer16}
\begin{equation}
t_{\rm dip} \approx e^{\frac{\pi^2 c}{8\beta}}\,,
\end{equation}
which is exponentially large\footnote{Note that there may be an ${\cal O}(1)$ factor multiplying the exponent, depending on whether one estimates the dip time due to only the vacuum character (as we did here), or whether one includes further matter contributions that can decay more slowly.} in the central charge $c$. Similarly the time after which the linear ramp joins the plateau is parametrically large in $c$ compared to the dip time.

Completely analogous behavior is found in the SYK model \cite{Cotler16}, believed to be dual to a black hole in AdS$_2$.

We therefore conclude that the RMT computation of the spectral form factor gives an excellent analytical model of the behavior in holographic theories, and may therefore possibly serve as an excellent toy model to describe information loss and restoration in black hole physics.\subsubsection{Bound on Fidelity Decay}
Let us now turn to the bound on the fidelity decay in terms of energy fluctuations in the thermal ensemble (\ref{Fcv}). Let us thus calculate the specific heat of the holographic dual of the thermofield double state, given, as we argued, by the eternal black hole \cite{Maldacena:2001kr}. Our interest is mainly in the black hole with compact horizon, since this corresponds to the case with a finite Hilbert space and thus a discrete spectrum.

In spacetime dimension $d$ one finds for the partition function
\begin{equation}
{\rm log}Z(\beta) = k_dc\beta\left( r_+^{d} - r_+^{d-2}\ell^2 \right)
\end{equation}
where we work in a scheme which sets the free energy of AdS$_{d+1}$ without a black hole ($r_+ = 0$) to zero. Again, $c$ is the central charge, and $k_d$ is a dimension-dependent constant, which can be found for example in \cite{EmparanCliff}. The radius of the black hole is set by $r_+$, which is related to the inverse temperature $\beta$ via the relation,
\begin{equation}
\beta = \frac{4\pi \ell^2 r_+}{dr_+^2 + (d-2)\ell^2}\,.
\end{equation}
 The special case of AdS$_3$ we considered above has $d=2$ and $r_+ = \frac{2\pi \ell^2 }{\beta}$. This allows us to determine the energy fluctuations $\Delta E$,
which together with Eq. (\ref{Fcv}) bound the fidelity decay. We start with the AdS$_3$ case, where we find
\begin{equation}
S(\beta,t) \ge \exp\left( -  \sqrt{\frac{2c}{3\beta^3}}\,\,2\pi t \right),
\end{equation}
in accordance with the value for $\Delta E$ appearing in the holographic Zeno time (\ref{eq.HoloZenoTime}).
For the general $d$-dimensional case we find
\begin{equation}\label{eq.enFlucAdS}
\Delta E^2 = \frac{8 (d-1)c\ell^2 \pi^2  k_d  r_+^d}{\beta^2(2\pi r_+ + (2-d)\beta)}\propto c
\end{equation}
leading to a bound of the form $S(\beta,t) \ge \exp\left( - k'_d \sqrt{c} t\right)$ for a $d$ dimensional constant $k_d'$ which follows upon taking the square root of (\ref{eq.enFlucAdS}).
\section{Conclusions and discussion}\label{sectionDis}
In this paper we have established that the spectral form factor $g(\beta, t)$ can be reformulated in terms of a decay amplitude of a certain pure state, namely the thermofield double state. The time evolution of $g(\beta, t)$ is thus mapped to the decay amplitude, as a function of time, of the fidelity of the TDS and the associated decay probability $S(\beta,t)$. 

Our results fall into two main categories. Firstly we make use of quantum-speed-limit and unitarity type bounds on the evolution of the fidelity of the TDS, leading to a bound on the early time gaussian decay, as well as a sub-exponential bound for the subsequent decay. The very late-time behavior has been shown to be governed by the purity of the TDS and is thus generally non-vanishing in theories with a discrete spectrum. This is to be contrasted with the complete decay of the spectral form factor when the spectrum is continuous. In theories with a discrete spectrum continued decay (perhaps for an intermediate time scale) should be interpreted as a sign of information loss, and so our results here help to sharpen this discussion, in particular in the context of holographic duality.

The second main category of results concerns the behavior of $g(\beta, t)$ in various models that we control analytically. In particular we have shown that the exact result for the spectral form factor for all times can be computed in the generalized unitary ensemble of random matrices using the method of orthogonal polynomials. The full analytical result not only matches earlier numerical studies \cite{Cotler16}, but is also remarkably similar (at all times) to the spectral form factor in various holographic models, such as the SYK model  \cite{Cotler16} as well as the BTZ black hole \cite{Dyer16}. This contrasts with the case of integrable models, as demonstrated in sections \ref{sec.SHOxp} and \ref{eq.CSmodel}. Our results in the GUE should thus serve as an analytical toy model of the study of information loss and retrieval in black holes. It will particularly interesting to study the relation of information restoration to non-perturbative corrections with respect to the large$-N$ approximation \cite{Anous:2016kss,Fitzpatrick:2016ive}.

{\it Acknowlegments.---} 
We would like to thank Ethan Dyer, Yoan Emery, Gautam Mandal, Marcos Mari\~no, Lea F. Santos and Benjamin Withers for helpful conversations.
Funding support from the John Templeton foundation and UMass Boston (project P20150000029279)  is  acknowledged. JMV is supported by Ministerio de Econom\'{i}a y Competitividad FIS2015-69512-R and Programa de Excelencia de la Fundaci\'{o}n S\'{e}neca 19882/GERM/15. JS acknowledges funding from the Fonds National Suisse pour la recherche scientifique and SwissMAP, the Swiss National Center for Competence in Research ``The Mathematics of Physics".
\appendix\label{app}

\section{Bhattacharyya bound for arbitrary initial states}\label{GBB}

The bound derived in \cite{Bhattacharyya83} holds for an initial pure state that undergoes unitary evolution. The fidelity between the initial and time-dependent state equal the survival probability  defined as $S(t)={\cal F}[|\psi(0)\ra,|\psi(t)\ra]=|\la \psi(0)|\psi(t)\ra|^2\in[0,1]$. In this appendix, we present a generalization that holds for arbitrary initial states, including mixed states described by an arbitrary density matrix. 
It is the convenient to use $S(t)={\cal F}\left( \rho_0,\rho_t\right)=\left[{\tr \sqrt{\sqrt{\rho_0}\,\rho_t\, \sqrt{\rho_0}}}\, \right]^2$ to quantify the notion of distinguishability between the initial and time-dependent density matrix, $\rho_0$ and $\rho_t$.
We note that the rate of change of $\mathcal{L}\left(\rho_0,\rho_t\right)=\cos^{-1}\left[{\tr \sqrt{\sqrt{\rho_0}\,\rho_t\, \sqrt{\rho_0}}}\, \right]$,
\beqa
\frac{d}{dt}\mathcal{L}\left(\rho_0,\rho_t\right)\leq \bigg|\frac{d}{dt}\mathcal{L}\left(\rho_0,\rho_t\right)\Bigg|=\frac{|\dot{S}|}{2\sqrt{(1-S)S}}.
\eeqa
We further use Uhlmann's bound \cite{Uhlmann92}
\beqa
\frac{d}{dt}\mathcal{L}\left(\rho_0,\rho_t\right)\leq\Delta E, 
\eeqa
to find
\beqa
\label{dcos}
\frac{|\dot{S}|}{2\sqrt{(1-S)S}}\leq\Delta E.
\eeqa

Equation (\ref{dcos}) amounts to
\beqa
\frac{d}{dt} \arccos\sqrt{S}\leq \Delta E,
\eeqa
that upon integration leads to the Mandelstam-Tamm (MT) quantum speed limit 
\beqa
t\geq t_{ MT}\equiv \frac{\mathcal{L}\left(\rho_0,\rho_t\right)}{\Delta E},
\eeqa
first obtained by Uhlmann for mixed states \cite{Uhlmann92}.
Thus, for times $0\leq t\leq \pi/(2\Delta E)$, 
\beqa
S(t)\geq \cos^2(\Delta E t).
\eeqa

We further note that $\sqrt{(1-S)S}\leq S$ for $S \geq 1/2$.
Then, as $S=|S |$ and $\int |\dot{S}/{S} | dt \geq |\int \dot{{S}}/{S} dt|=|\ln S|=-\ln S$, it follows that
\beqa
\label{B2}
S(t)\geq \exp(-2\Delta E t),
\eeqa
that for times $t\geq 0$ when  $S(t)\geq 1/2$, until the half lifetime $t_{h}$ when the fidelity first reaches  $S(t_{h})=1/2$. The lifetime $t_h$ can exceed $\pi/(2\Delta E)$. Equation (\ref{B2}) is the generalized Bhattacharyya bound for an arbitrary initial state.

\section{Complete decay for systems with continuous spectrum}\label{AppendixFock}
\label{sec.CompleteDecay}
Whenever the spectrum of the Hamiltonian generating the time evolution is continuous, the vanishing survival probability of a  thermofield double state at infinitely long time of evolution follows form the analytic properties of the density of states.
To show this, it is convenient to rewrite the survival probability as a Fourier transform.
Let $\hat{H}$ be the driving Hamiltonian for $t\geq 0$ with a continuous spectrum,
\beqa
\hat{H}|E\ra=E|E\ra, \quad E>E_0, 
\eeqa 
where we assume  the existence of a ground state energy $E_0$, so that the Hamiltonian is bounded from below.
The resolution of the identity in terms of the scattering states reads
\beqa
\int_{E_0}^\infty dE|E\ra\la E|=1.
\eeqa
Using it, the survival probability can be written as
\beqa
S(t)&=&\left|\la\psi(0)|e^{-i\hat{H}t}|\psi(0)\ra\right|^2,\\
&=&\left|\int_{E_0}^\infty dE\la\psi(0)|e^{-i\hat{H}t}|E\ra\la E|\psi(0)\ra  \right|^2,\\
&=&\left|\int_{E_0}^\infty dE \rho(E)e^{-iEt}\right|^2,
\eeqa
where we have defined the energy distribution of the initial state as the probability of finding $|\psi(0)\ra$ in one of the scattering states $|E\ra$,
\beqa
\rho(E)\equiv |\la E|\psi(0)\ra|^2.
\eeqa
This is a function with support on the domain $[E_0,\infty)$. 
In order to write the survival probability as a Fourier transform, we further define
\beqa
\tilde{\rho}(E)=\rho(E)\Theta(E-E_0),
\eeqa
with the full real line as a domain. In terms of it,
 \beqa
S(t)&=&\left|\int_{\mathbb{R}} dE \tilde{\rho}(E)e^{-iEt}\right|^2.
\eeqa
That $S(t)$ vanishes at infinity as a consequence of the Rieman-Lebesgue lemma \cite{FK47}.
This is a useful result  in many studies of quantum dynamics so let us state it in detail.
We recall the definition of an $L^p$-function $f(z)$,
\beqa
\| f\|_p \equiv \left(\int |f(z)|^p dz\right)^{\frac{1}{p}}<\infty.
\eeqa 
The Riemann-Lebesgue lemma states that if the Lebesgue integral of a function $|f|$ is finite then  
its Fourier transform vanishes as the conjugate variable tends to infinity.
In other words, let  $f$ be a $L^1$-function, then
\beqa
\hat{f}(t)\equiv \int_{\mathbb{R}} f(z)e^{-izt}\rightarrow 0 \quad {\rm as } \quad t\rightarrow \infty.
\eeqa
It is not difficult to show that $\tilde{\rho}(E)$ is a $L^1$-function
\beqa
|\tilde{\rho}(E)|_1&=&\int_{\mathbb{R}} dE|\tilde{\rho}(E)|,\\
&=&\int_{E_0}^\infty dE\la\psi(0)|E\ra\la E|\psi(0)\ra =1,
\eeqa
whence it follows that $S(t)\rightarrow 0$ as $t\rightarrow\infty$.

\section{Exact Survival Amplitude in GUE}\label{sec.ExactAmplitude}
Here we supply the details in the derivation of the survival probability $S(\beta,t)$ in the GUE random matrix ensemble in section \ref{sec.ExaFi} of the main text.
\subsection{Integrals of Hermite polynomials}
In what follows we shall find it necessary to evaluate integrals of the type
\beqa\label{eq.MasterIntegral}
{\cal I}_{n}(\nu) &:=& \int_{-\infty}^\infty dE e^{-\nu E - E^2}H_n(E)\,,\\
{\cal I}_{nm}(\nu) &:=& \int_{-\infty}^\infty dE e^{-\nu E - E^2}H_n(E) H_m(E)\,,\label{eq.MasterIntegral2}
\eeqa
over one or two Hermite polynomials.
To this end, the main tool we use is the identity
\beqa\label{eq.HermiteIdentity}
\int dE e^{-E^2}H_n(E) f(E) = \int dE e^{-E^2}D^n f(E)\,,
\eeqa
which follows from the usual definition of the Hermite polynomials\,,
\beqa
e^{-E^2}H_n(E) = (-D)^n e^{-E^2}\,,
\eeqa
after $n$ integrations by part. Here and below we use the notation $D := d/dE$. The integrands must decay sufficiently fast at infinity in order to justify neglecting the boundary terms. This is always the case for the integrals of interest here.

We now use this idea to evaluate the first integral appearing in (\ref{eq.MasterIntegral}). This reduces to an expression of the form (\ref{eq.HermiteIdentity}) for the choice 
$f(E) = e^{-\nu E}$. The resulting Gaussian integral is trivial, with the result
\beq\label{eq.OneHermiteIntegral}
{\cal I}_n (\nu) = \sqrt{\pi} \left( - \nu \right)^n e^{\frac{\nu^2}{4}}\,.
\eeq

We now proceed to evaluating the second integral,  (\ref{eq.MasterIntegral2}), appearing above. In this case one obtains by successive integration by parts (or simply by using the identity (\ref{eq.HermiteIdentity}) above) 
\beqa\label{eq.HermiteIntegralTrick}
{\cal I}_{nm}(\nu) = \int dE e^{-E^2}D^nf_m(E)
\eeqa
for 
\beqa
f_m(E) = e^{-\nu E}H_m(E)\,.
\eeqa
We now notice that we can compute this derivative using the binomial expansion
\beqa
D^n \left( e^{-\nu E}H_m(E) \right) &=& \sum_{k=0}^n  {}^n C_k D^k (e^{-\nu E})D^{n-k}H_m\,,\nonumber\\
&=& \sum_{k=0}^n {}^n C_ke^{-\nu E}  (-\nu)^kD^{n-k}H_m,\nonumber\\
\eeqa
where ${}^nC_k$ is the usual binomial coefficient. Furthermore, we have
\beqa
D^{n-k}H_m(E) = 2^{n-k}\frac{m!}{p!}H_{p}(E)\,,
\eeqa
with $p=m+k-n$, following standard properties of the Hermite polynomials. We have thus reduced the integral over two Hermite polynomials into a sum of integrals over single Hermite polynomials (times coefficients). For these we can use our previous result (\ref{eq.OneHermiteIntegral}) to find
\beqa\label{eq.MasterIntTwoResult}
{\cal I}_{nm} &=&\sqrt{\pi}e^{\frac{\nu^2}{4}}2^n \left( - \nu \right)^{m-n}\sum_{k=0}^n \frac{{}^n C_k \left(-\nu \right)^{2k}}{2^k (m+k-n)!},\nonumber\\
&=& \frac{\sqrt{\pi}2^nm!}{(m-n)!} \left( - \nu \right)^{m-n} {}_1F_1\left[  -n, 1 + m - n, -\tfrac{\nu^2}{2}\right],\nonumber\\
\eeqa
where $ {}_1F_1$ is a confluent hypergeometric function defined by the finite power series expansion in the line above. Let us note in passing that the diagonal case $m=n$ reduces to a simple Laguerre polynomial,
\beq\label{eq.DiagonalTwoHermiteIntegral}
{\cal I}_{nn}=\sqrt{\pi}2^n n! L_n \left( -\tfrac{\nu^2}{2}\right)\,.
\eeq

Armed with the master integrals we are now in a position to compute the spectral form factor quoted in the bulk of the paper, including the disconnected contributions, which reduce to a product of two (GUE averaged) partition function.

\subsection{Spectral density}
Using the exact eigenvalue density $\rho(E)$ as defined in (\ref{rhoGUE1}), we would like to compute the sum
\beqa
\cI (\nu) =   \sum_{n=0}^{N-1}c_n^2 \int_\mathbb{R} dE e^{-E^2-\nu E}H_n^2(E)
\eeqa
for the coefficients
\beqa\label{eq.HermiteCoeffs}
c_n^2 = \frac{1}{2^n n! \sqrt{\pi}}\,.
\eeqa
From this, the survival amplitude follows as
\beqa
A(\beta,t) = \frac{1}{Z}\cI(\nu)\,,
\eeqa
where $\nu = \beta + it$ and $Z = \cI(\nu)\Bigr|_{\nu = \beta}$. It is not hard to recognize that all coefficients conspire precisely to give a straight sum over Laguerre polynomials
\beqa
\cI (\nu)  &=& e^{\frac{\nu^2}{4}}\sum_{n=0}^{N-1} L_n \left( -\tfrac{\nu^2}{2}\right),\nonumber\\
&=&e^{\frac{\nu^2}{4}}L_{N-1}^1 \left( -\tfrac{\nu^2}{2}\right)\,.
\eeqa
In the last line we have reexpressed the sum in terms of a single associated Laguerre polynomial, recovering the expression in the main body of our paper, Eq. (\ref{eq.ExactGUEfidelityDecay}). We now turn to the two-level correlation function.

\subsection{Two-level correlation}

In the case of the connected two-level correlation (\ref{eq.TwoLevelConnectedCorrelation}), 
we start by rewriting the definition (\ref{eq.TwoLevelConnectedCorrelation}) together with the two-level cluster function (\ref{eq.TwoLevelCorrelationHermite}) as 
\beq\label{eq.IntegralFactor}
g_c(\beta, t) =\sum_{n,m=0}^{N-1}c_n^2 c_m^2 {\cal I}_{nm} (\nu){\cal I}_{mn} (\bar \nu)\,,
\eeq
which follows from noticing that
\beq
\left( \sum_{n=0}^{N-1}\varphi_n(E_1)\varphi_n(E_2)  \right)^2 = \sum_{m,n=0}^{N-1}\varphi_n(E_1)\varphi_m(E_2)\varphi_m(E_1)\varphi_n(E_2)
\eeq
so that the integral (\ref{eq.TwoLevelConnectedCorrelation}) factorized as indicated in (\ref{eq.IntegralFactor}). We then combine the second master integral (\ref{eq.MasterIntTwoResult}) with the coefficients (\ref{eq.HermiteCoeffs}) to arrive at the expression quoted in the bulk of the paper (\ref{eq.connectedFormFactorGUE}).

\subsection{Map to Wilson loop}\label{sec.MapToWilson}

It is amusing to note that the computation of the survival amplitude can be mapped exactly to the computation of a Wilson loop in the Gaussian matrix model. To be precise, the partition function, satisfies
\beq
\left\langle Z(\nu) \right\rangle = \left\langle W_C(2 g_s) \right\rangle
\eeq
while the spectral form factor satisfies
\beq
\left\langle Z(\nu) Z(\bar\nu)\right\rangle = \left\langle W_C(2 g_s) W_C(2 \overline{g}_s) \right\rangle
\eeq
We shall now prove these relations. Let us start with the partition function
\begin{widetext}
\beqa
\int\sum_n\left\langle \delta \left(E - E_n \right)\right\rangle_{\rm GUE}  e^{-\nu E} dE &=&\int\left( \int \prod_i dE_i \Delta^2 (E) \sum_n \delta \left(E - E_n \right) e^{-\sum_i E_i^2}\right) e^{-\nu E}dE\,,\nonumber\\
&=& \int \prod_i dE_i \Delta^2(E) e^{-\sum_i E_i^2}\sum_n e^{-\nu E_n} \,,
\eeqa
where in the last line we have performed the integral over $E$ with the help of the delta function insertions. We have written the measure of the GUE in terms of an intergral over eigenvalues $E_i$ with the Jacobian factor $\Delta^2(E)$ given by the Vandermonde determinant
\beq
\Delta(E) = \prod_{i<j}|E_i - E_j|\,.
\eeq
But now we can relate the latter expression to the expectation value of the insertion of a single Wilson loop operator, given by $W = {\rm Tr}e^M$, in a Gaussian matrix model over Hermitian matrices $M$. Explicitly, 
\beq
 \int \prod_i dE_i \Delta^2(E)\sum_n e^{-\nu E_n}  e^{-\sum_i E_i^2} = \frac{1}{\Omega_M}\int dM {\rm Tr}e^{-\nu M}e^{-{\rm Tr}M^2} := \langle W_C(\nu)\rangle\,,
\eeq
where $\Omega_M$ is a normalisation factor, whose precise form does not concern us here and $dM$ indicates the Haar measure over Hermitian matrices.
In order to connect to the standard presentation of the matrix model, we should identify the parameter $\nu$ formally with the coupling $g_s$, via
\beq\label{eq.GaugeCoupling}
\nu^2 = 2g_s\,,
\eeq
which appears in the standard weight of the Gaussian matrix model as $e^{-\frac{1}{2g_s}{\rm Tr}M^2}$. Having understood the map of the GUE average of the partition function, it is then a matter of applying the same sort of manipulations to show that
\beq
\left\langle  Z(\nu)Z(\bar\nu)\right\rangle = \int dE dE'\left\langle \sum_n \delta(E-E_n) \sum_m \delta(E'-E_m)\right\rangle_{\rm GUE} e^{-\nu E - \bar \nu E'} = \langle W_C(\nu)W_C(\bar\nu)\rangle\,,
\eeq
establishing, as claimed, that the spectral form factor maps to a correlation function of two Wilson loops.
Here the same formal identification of the gauge coupling (\ref{eq.GaugeCoupling}) is again  implied.
\end{widetext}

\end{document}